\documentclass[twocolumn]{revtex4}

\usepackage{color} 
\usepackage{graphicx} 
\usepackage{amsmath,amssymb}

\begin{document}

\title{Rethinking hydrogen-bond kinetics}

\author{Diego Prada-Gracia}
\author{Francesco Rao} 
\affiliation{Freiburg Institute for Advanced Studies,
School of Soft Matter Research, Albertstrasse 19, 79104 Freiburg im Breisgau,
Germany}

\maketitle

\textbf{Abstract.} At the fundamental level, our understanding of water hydrogen-bond dynamics has been largely built on the detailed analysis of classical molecular simulations. The latter served to develop a plethora of hydrogen bond definitions based on different properties, including geometrical distances \cite{Buch1992Growth,Luzar1996Effect,Kumar2007Hydrogen}, topology \cite{Hammerich2008Alternative,Smith2005Unified,Henchman2010Topological} and energetics \cite{Geiger1979Aspects,Jorgensen1980MonteCarlo}. Notwithstanding, no real consensus emerged from these approaches, making the development of a consistent and reliable definition elusive \cite{Prada2013Quest}. In this contribution, a framework to study hydrogen bonds in liquid water based purely on kinetics is presented. This approach makes use of the analysis of commitment probabilities without relying on arbitrarily chosen order parameters and cutoffs. Our results provide evidence for a self-consistent description, resulting in a clear multi-exponential behavior of the kinetics.

The last decade represented a real \emph{golden age} for the
characterization of states in complex molecular systems. With
the emergence of kinetics-based models, it is now possible to
characterize with unprecedented detail the free-energy landscape of a
molecule without the use of pre-chosen order parameters. To do so,
projection-free approaches make use of the kinetics to define states
and transition saddles. This is possible by working on a discrete
representation of the accessible space in terms of microstates and the
transitions among them. Within this framework, two molecular configurations are
considered to belong to the same free-energy basin (i.e., same
\emph{state}) if they interconvert rapidly into each other. Such
information can be inferred by various techniques based on transition
networks, including complex network analysis
\cite{Rao2004Protein,Gfeller2007Complex,Rao2010Protein}, Markov state models
\cite{Chodera2007Automatic,Prinz2011Optimal} or local fluctuations
analysis
\cite{Baba2007Construction,Berezovska2012Accounting}. However, the
development of a reliable microstate definition for the characterization
of hydrogen bonds is hard \cite{Rao2010Structural, Prada2012Towards},
limiting the use of transition networks for this problem.

Conceptually speaking, networks-based methods make implicit use of \emph{commitment probabilities} \cite{Rao2005Estimation} for the characterization of the dynamics. While this technique alone cannot cope with multi state problems, it represents a viable approach for two state systems. In this contribution, we present a framework making use of commitment probabilities for the characterization of hydrogen bonds. Our analysis is based on a molecular dynamics simulation of 1024 TIP4P-Ew water molecules \cite{Horn2004Development} at 300K and ambient pressure with a saving frequency of 4 fs. The approach works as follows. First, choose a water molecule to analyze. This molecule is called the \emph{target}. Second, for every other water molecule calculate the time series of the ranking number with respect to the target. The ranking number coordinate is equal to 1 when a molecule has the shortest inter-oxygen distance with respect to the target, 2 if it is the second closest and so on (see top of Fig.~1 for an example time series). Third, calculate the commitment probability $P_C$. Two molecules will be considered to be bound, unbound or close to the transition state according to the value of $P_C$. 

\begin{figure}[t!]
  \includegraphics[width=80mm]{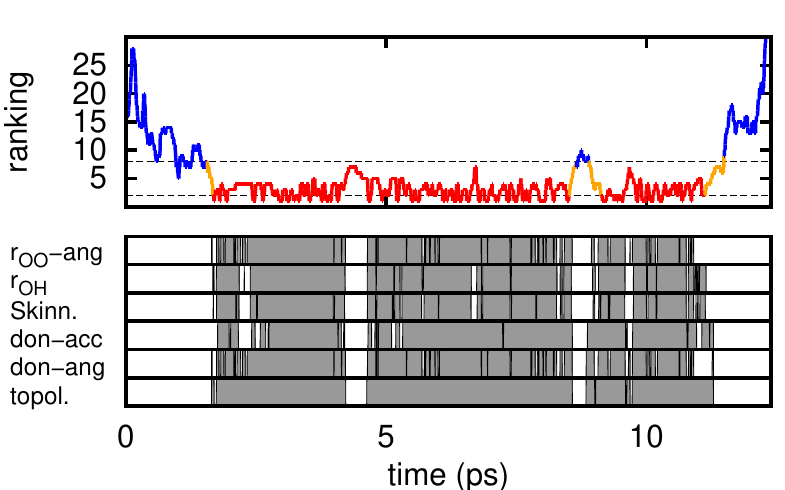}
  \caption{Hydrogen bond time series. (Top) Time series of the ranking number of a
    water molecule with respect to another (the target). Bound, unbound and transition regions according to commitment probabilities are shown in red,
    blue and orange, respectively. The sink and source-thresholds are shown as dashed lines. (Bottom) Gray filled boxes represent hydrogen bond detection according to the following definition: r$\mathrm{_{OH}}$ \cite{Buch1992Growth},
    r$\mathrm{_{OO}}$-angle \cite{Luzar1996Effect}, Skinner's
    \cite{Kumar2007Hydrogen}, donor-acceptor
    \cite{Hammerich2008Alternative}, donor-angle
    \cite{Smith2005Unified} and topological
    \cite{Henchman2010Topological}. }
  \label{fig:segment}
\end{figure}

\begin{figure}[t!]
  \includegraphics[width=80mm]{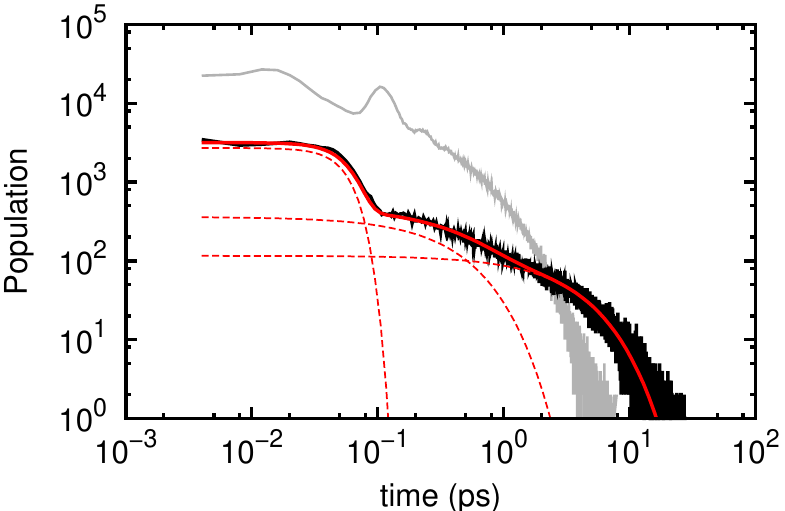}
  \caption{Hydrogen bond lifetime distributions. Results obtained from commitment probabilities and Skinner's hydrogen bond definition are shown in black and gray, respectively. Individual fitting functions and the total fit are shown as red dashed and red solid lines. The fit function is: $a\exp{[-(x/\tau_a)^{\gamma}]} + 
    b\exp{(-x/\tau_b)} + c\exp{(-x/\tau_c)}$, where $a$, $b$ and $c$ are $2.7\cdot 10^{3}$, $3.6\cdot 10^{2}$ and $1.15\cdot 10^{2}$ respectively; $\tau_a$, $\tau_b$
    and $\tau_c$ are equal to 0.06, 0.4 and 3.45 ps; and $\gamma=2.98$.}
  \label{fig:kinetics}
\end{figure}

For the case of a perfect order parameter, the position of the barrier is defined by a specific value of the coordinate. This is not the case for the ranking number where values around 4 or 5 cannot be always unambiguously assigned to the bound or the unbound states. Hence, to perform a commitment probability analysis a \emph{sink} and \emph{source} thresholds need to be set \cite{Rao2005Estimation}. They respectively represent the border of the bound and unbound states once the barrier region has passed. Note that no assumption on the exact position of the barrier is made. For the sink, a ranking number of 1 or 2 can be chosen,  while for the source the decision is not straightforward.  The source-threshold represents a situation where a molecule is stably unbound with no fast re-crossings to the bound state (i.e. quickly re-binding molecules that didn't relax to the unbound state). For example, a ranking around 5 will not work, since several fluctuations to this rank exist from both the bound and unbound states. On the other hand, choosing a rank which is too high will be too conservative, potentially missing a considerable part of the unbound state. To avoid an arbitrary decision, the flux (i.e. the total number of transitions) from the bound to the unbound states was calculated for different values of the source-thresholds. For small values of the source-threshold the flux tends to be high due to the large number of re-crossings. As the value of the threshold increases, the flux monotonically decreases with a kink at around ranking number eight (not shown). The kink indicates that for that number the flux is not affected  by re-crossings anymore. For larger values of the ranking, the flux still slightly decreases because the volume of the unbound region is progressively reduced. Given this reasoning, an arbitrary decision was avoided taking as sink and source thresholds ranking numbers 2 and 8, respectively (horizontal dashed bars in Fig.~1, top).

The commitment probability $P_C$ is then calculated along the ranking number time series. For any given time, $P_C=1$ if the sink is the first threshold found when the history is tracked both towards the past (backward) and the future (forward) (red in Fig.~1, top); if in both cases the source-threshold is met first, then the molecule is unbound ($P_C=0$, blue); otherwise, the molecule belongs to a transition region going from source to sink or vice-versa ($P_C=0.5$, orange). Using this approach the relation between two water molecules is solely based on kinetics, taken as only assumption that two  mutually interacting molecules will also be close in space (this is the reason to use a distance-based coordinate). No assumptions on the energetics, geometry or topology are done. Moreover, an arbitrary decision on the exact position of the barrier is avoided: a problem that has always affected cutoff-based definitions \cite{Prada2013Quest}.

In the example time series of Fig.~\ref{fig:segment}, the ranking number stays always very high when two molecules do not interact ($t<1.5$ and $t>11.5$). On the other hand, when the interaction is formed the ranking number rarely exceeds the value of 4 or 5. Human eye can easily recognize the bound region without being distracted by the sporadic fluctuations. This is not the case for classical hydrogen bond definitions where fast fluctuations can momentarily disrupt the interaction being followed by a quick re-binding (re-crossings). Predictions from six among the most commonly used classical hydrogen bond definitions are shown at the bottom of Fig.~\ref{fig:segment}. Some observations can be done: first, as already investigated by us in a recent contribution \cite{Prada2013Quest}, despite qualitative agreement no clear consensus emerges among the definitions. Second, several re-crossings are observed (intermitting short white spaces). This is not the case for the regions with $P_C=1$ (in red) where contiguous intervals, with no evidence of re-crossings, are found. In particular, the two bound events are separated by a hydrogen bond switch (same water pair interacting with a different donated hydrogen), while the other large fluctuation at around 4.5 ps, although similar in amplitude, represents only a partial stretch of the same hydrogen bond (confirmed by visual inspection of the molecular trajectory).

The high quality of the characterization of the kinetics is demonstrated by the shape of the lifetime distribution (black line in Fig.~\ref{fig:kinetics}). The curve can be well fit by a sum of exponentials (showed in red). This is not the case for more conventional hydrogen bond definitions as it is well known in the literature \cite{Sciortino1990Lifetime,Luzar1996Hydrogen,Prada2013Quest}. Typical behavior presents both an oscillating section at fast times and a non-exponential decay at large times (see gray line for an example). The exponential behavior obtained with the commitment probability provides further points of discussion. First, no oscillating behavior is observed at fast time scales. Moreover, the number of events in this region is smaller by one order of magnitude with respect to the conventional distribution, indicating a dramatic reduction of re-crossings. Second, an increased number of events at large times reveals the existence of longer-lived hydrogen bonds lasting more than 10 ps: a somewhat unexpected result for liquid water at ambient conditions. Third, the entire curve can be fitted simultaneously by a combination of a compressed exponential and two exponential functions. Contrary to the previously observed non-exponential behavior, the appearance of a double exponential provides a more rational way to interpret the data in terms of two independent processes (that can be qualitatively interpreted as distorted and perfectly aligned hydrogen bonds, respectively). Finally, the compressed exponential behavior at fast time scales is due to repulsions when two molecules face their negative charges one in front of the other. This occurs because the use of the ranking time series does not restrict ourselves to hydrogen bonds but to all type of inter molecular interactions, including repulsions.

Notwithstanding this approach cannot be used straightforwardly as more conventional hydrogen bond definitions, our results provide a solid starting point for the development of more accurate, kinetics-based, characterization of hydrogen bonds.


\end{document}